\documentstyle[11pt]{article}
\makeatletter%
\def\nottoobig#1{{\hbox{$\left#1\vcenter to1.111\ht\strutbox{}\right.\n@space$}}}
\makeatother%

\makeatother%

\makeatletter%

\newcount\hour  \newcount\minutes  \hour=\time  \divide\hour by 60
\minutes=\hour  \multiply\minutes by -60  \advance\minutes by \time
\def\mmmddyyyy{\ifcase\month\or Jan\or Feb\or Mar\or Apr\or May\or Jun\or Jul\or
  Aug\or Sep\or Oct\or Nov\or Dec\fi \space\number\day, \number\year}
\def\hhmm{\ifnum\hour<10 0\fi\number\hour :%
  \ifnum\minutes<10 0\fi\number\minutes}
\def\Draft{{\it Draft of \mmmddyyyy}}

\def\ps@jtsheadings{%
\def\@oddhead{\it\rightmark\hfil\rm\thepage}%
\def\@oddfoot{\hfil\Draft}%
\if@twoside%
\def\@evenhead{\rm\thepage\hfil\it\leftmark}%
\def\@evenfoot{\Draft\hfil}%
\else
\let\@evenhead\@oddhead%
\let\@evenfoot\@oddfoot%
\fi%
}
\def\ps@jtsplain{%
\def\@oddhead{\hfil\Draft}%
\def\@oddfoot{\hfil\rm\thepage\hfil}%
\let\@evenfoot\@oddfoot%
\if@twoside \def\@evenhead{\Draft\hfil} \else \let\@evenhead\@oddhead \fi
}

\def\chaptermark#1{\markboth{\thechapter.\ #1}{\thechapter.\ #1}}%
\def\sectionmark#1{\markright{\thesection.\ #1}}

\def\section{\@startsection {section}{1}{\z@}
    {3.5ex plus1ex minus.2ex}{2.3ex plus.2ex}{\Large\bf}}
\def\subsection{\@startsection{subsection}{2}{\z@}
    {3.25ex plus1ex minus.2ex}{1.5ex plus.2ex}{\large\bf}}
\def\subsubsection{\@startsection{subsubsection}{3}{\z@}
    {3.25ex plus1ex minus.2ex}{1.5ex plus.2ex}{\normalsize\bf}}
\def\paragraph{\@startsection{paragraph}{4}{\z@}
    {3.25ex plus1ex minus.2ex}{1em}{\normalsize\bf}}
\def\subparagraph{\@startsection{subparagraph}{4}{\parindent}
    {3.25ex plus1ex minus.2ex}{1em}{\normalsize\bf}}

\makeatother%

\makeatletter \@beginparpenalty=10000 \makeatother

\def\underl#1 {\leavevmode\let\first=\relax\underli #1 }
\def\underli#1 {\ifx&#1\let\next=\relax\unskip
                \else\let\next=\underli\first\ulinebox{#1}\fi\let\first=\undersp\next}
\def\undersp{\penalty50\ulinebox{\space}\penalty50}
\def\ulinebox#1{\vtop{\hbox{\strut#1}\hrule}}%
\def\unice#1 {\underl #1 & }
\def\desclabel#1{\bf #1\hfil}
\def\desc{\list{}{%
\labelwidth=\leftmargin
\advance \labelwidth by -\labelsep
\let \makelabel=\desclabel}}

\makeatletter %

\newcommand{\implies}{\:\Rightarrow\:}

\newlength{\filength}
\newsavebox{\gcbox}
\sbox{\gcbox}{\framebox[\filength]{\rule{0ex}{2ex}}}

\newlength{\leftjustindent}
\newlength{\@leftjustindent}
\def\leftjust{\let\\\@leftjustcr\let\end\@endleftjust
  \addtolength{\@leftjustindent}{\leftjustindent}
  \vcenter\bgroup
  \halign\bgroup
    \hbox to\displaywidth{
      \rule{\@leftjustindent}{0ex}$\displaystyle##$\hfill
      }\crcr
}
\def\endleftjust{\crcr\egroup\egroup\endgroup}
\def\@endleftjust#1{\crcr\egroup\egroup\@checkend{#1}\endgroup}
\def\@leftjustcr{\crcr}

\newenvironment{proof}{\Proof}{\qed}
\newtheorem{theorem}{Theorem}[section]
\newtheorem{corollary}[theorem]{Corollary}
\newtheorem{definition}[theorem]{Definition}

\def\Proof{\par\medskip\noindent{\bf Proof}\\\mbox{\rule{\parindent}{0in}}}
\newcommand{\qedblob}{\mbox{\rule[-1.5pt]{5pt}{10.5pt}}}
\def\literalqed{{\ \nolinebreak\hfill\mbox{\qedblob\quad}}}

\def\qed{\literalqed}

\newtheorem{lemma}[theorem]{Lemma}

\newcommand{\singlespacing}{\let\CS=
\@currsize\renewcommand{\baselinestretch}{1}\tiny\CS}
\newcommand{\singlespacingplus}{\let\CS=
\@currsize\renewcommand{\baselinestretch}{1.25}\tiny\CS}
\newcommand{\doublespacing}{\let\CS=
\@currsize\renewcommand{\baselinestretch}{1.75}\tiny\CS}
\newcommand{\draftspacing}{\let\CS=
\@currsize\renewcommand{\baselinestretch}{2.0}\tiny\CS}

\makeatother%

\mathcode`\0="0030      %
\mathcode`\1="0031
\mathcode`\2="0032
\mathcode`\3="0033
\mathcode`\4="0034
\mathcode`\5="0035
\mathcode`\6="0036
\mathcode`\7="0037
\mathcode`\8="0038
\mathcode`\9="0039
\makeatletter
\clubpenalty=\@highpenalty
\widowpenalty=\@highpenalty
\makeatother

\makeatletter
\newcommand{\niceonespacing}{\let\CS=\@currsize\renewcommand{\baselinestretch}{1.1}\tiny\CS}\newcommand{\nicetwospacing}{\let\CS=\@currsize\renewcommand{\baselinestretch}{1.2}\tiny\CS}
\newcommand{\nicethreespacing}{\let\CS=\@currsize\renewcommand{\baselinestretch}{1.3}\tiny\CS}
\newcommand{\singlespacingplusplus}{\let\CS=\@currsize\renewcommand{\baselinestretch}{1.35}\tiny\CS}
\newcommand{\nicefourspacing}{\let\CS=\@currsize\renewcommand{\baselinestretch}{1.4}\tiny\CS}
\newcommand{\nicefivespacing}{\let\CS=\@currsize\renewcommand{\baselinestretch}{1.5}\tiny\CS}
\newcommand{\nicesixpacing}{\let\CS=\@currsize\renewcommand{\baselinestretch}{1.6}\tiny\CS}
\makeatother

\makeatletter
\def\@cite#1#2{[#1\if@tempswa , #2\fi]}
\makeatother

\makeatletter
\def\@citex[#1]#2{\if@filesw\immediate\write\@auxout{\string\citation{#2}}\fi
  \def\@citea{}\@cite{\@for\@citeb:=#2\do
    {\@citea\def\@citea{,\linebreak[0]}\@ifundefined
       {b@\@citeb}{{\bf ?}\@warning
       {Citation `\@citeb' on page \thepage \space undefined}}%
\hbox{\csname b@\@citeb\endcsname}}}{#1}}
\makeatother

\makeatletter
\def\ps@thesis{\def\@oddhead{\hfil\rm\thepage\hfil}\def\@oddfoot{}\def\@evenhead{\hfil\rm\thepage\hfil}\def\@evenfoot{}\def\chaptermark##1{}\def\sectionmark##1{}}
\makeatother

\makeatletter
\def\foobarpt{\textfont\z@\tenrm 
  \scriptfont\z@\ninrm \scriptscriptfont\z@\sevrm
\textfont\@ne\tenmi \scriptfont\@ne\ninmi \scriptscriptfont\@ne\sevmi
\textfont\tw@\tensy \scriptfont\tw@\ninsy \scriptscriptfont\tw@\sevsy
\textfont\thr@@\tenex \scriptfont\thr@@\tenex \scriptscriptfont\thr@@\tenex
\def\unboldmath{\everymath{}\everydisplay{}\@nomath\unboldmath
          \textfont\@ne\tenmi 
          \textfont\tw@\tensy \textfont\lyfam\tenly
          \@boldfalse}\@boldfalse
\def\boldmath{\@ifundefined{tenmib}{\global\font\tenmib\@mbi\@magscale1\global
        \font\tensyb\@mbsy \@magscale1\global\font
         \tenlyb\@lasyb\@magscale1\relax\@addfontinfo\@xiipt
              {\def\boldmath{\everymath
                {\mit}\everydisplay{\mit}\@prtct\@nomathbold
                \textfont\@ne\tenmib \textfont\tw@\tensyb 
                \textfont\lyfam\tenlyb\@prtct\@boldtrue}}}{}\@xiipt\boldmath}%
\def\prm{\fam\z@\tenrm}%
\def\pit{\fam\itfam\tenit}\textfont\itfam\tenit \scriptfont\itfam\ninit
   \scriptscriptfont\itfam\sevit
\def\psl{\fam\slfam\tensl}\textfont\slfam\tensl 
     \scriptfont\slfam\tensl \scriptscriptfont\slfam\tensl
\def\pbf{\fam\bffam\tenbf}\textfont\bffam\tenbf 
   \scriptfont\bffam\ninbf \scriptscriptfont\bffam\ninbf 
\def\ptt{\fam\ttfam\tentt}\textfont\ttfam\tentt
   \scriptfont\ttfam\nintt \scriptscriptfont\ttfam\nintt 
\def\psf{\fam\sffam\tensf}\textfont\sffam\tensf
    \scriptfont\sffam\tensf \scriptscriptfont\sffam\tensf
\def\psc{\@getfont\psc\scfam\@xiipt{\@mcsc\@magscale1}}%
\def\ly{\fam\lyfam\tenly}\textfont\lyfam\tenly 
   \scriptfont\lyfam\ninly \scriptscriptfont\lyfam\sevly
 \@setstrut \rm}

\makeatother

\newcommand{\p}{{\rm P}}

\newcommand{\littlepitalic}{{\it p}}

\newcommand{\np}{{\rm NP}}

\newcommand{\sigmaj}{{{\rm \Sigma}_j^p}}
\newcommand{\sigmak}{{{\rm \Sigma}_{k}^p}}
\newcommand{\sigmaq}{{{\rm \Sigma}_{q}^p}}

\newcommand{\sigmakplusone}{{\rm \Sigma}_{k+1}^p}
\newcommand{\sigmai}{{\rm \Sigma}_i^p}

\newcommand{\pik}{{\rm \Pi}_k^p}
\newcommand{\pikplusone}{{\rm \Pi}_{k+1}^p}

\newcommand{\psigjk}{  {\p^{(\sigmaj , \sigmak)}}}

\newcommand{\psigkone}{ {\p^{\sigmak[1]}}}
\newcommand{\psigktwo}{ {\p^{\sigmak[2]}}}
\newcommand{\lsigk}{ {L_{\sigmak}}}
\newcommand{\psigkmtt}{  {\p^{\sigmak}_{m \hbox{-}{\rm tt}}}}
\newcommand{\psigkmponett}{  {\p^{\sigmak}_{m+1 \hbox{-}{\rm tt}}}}
\newcommand{\psigimk}{  {\p^{(\sigmai , \sigmak)}_{1,m \hbox{-}{\rm tt}}}}

\newcommand{\psigjmk}{  {\p^{(\sigmaj , \sigmak)}_{1,m \hbox{-}{\rm tt}}}}
\newcommand{\psigik}{  {\p^{(\sigmai , \sigmak)}}}
\newcommand{\psigjustnpk}{  {\p^{(\np , \sigmak)}}}
\newcommand{\psigij}{  {\p^{(\sigmai , \sigmaj)}}}

\newcommand{\psigione}{ {\p^{\sigmai[1]}}}
\newcommand{\psigipone}{{\p^{\Sigma_{i+1}^p[1]}}}

\newcommand{\lpsigi}{ {L_{\psigione}}}

\newcommand{\lpsigipone}{{L_{\p^{\Sigma_{i+1}^p[1]}}}}
\newcommand{\diffmsigk}{{\rm DIFF}_m(\sigmak)}

\newcommand{\diff}{{\rm DIFF}}
\newcommand{\codiffmsigk}{{\rm co}{\diffmsigk}}

\newcommand{\ldiffmsigk}{{L_{\diffmsigk}}}
\newcommand{\deltatilde}{\tilde{\Delta}}

\newcommand{\bpp}{{\rm BPP}}

\newcommand{\conp}{{\rm coNP}}

\newcommand{\pspace}{{\rm PSPACE}}

\newcommand{\npnp}{{\np^{\rm NP}}}
\newcommand{\conpnp}{{\conp^{\rm NP}}}

\newcommand{\poly}{{\rm poly}}
\newcommand{\ph}{{\rm PH}}

\def\pair#1{{{\langle\!\!~#1~\!\!\rangle}}}

\newcommand{\manyone}{\mbox{$\,\leq_{\rm m}^{{\littlepitalic}}$\,}}

\newcommand{\sigmastar}{\mbox{$\Sigma^\ast$}}

\newcommand{\calc}{{\cal C}}

\newcommand{\condition}{\,\nottoobig{|}\:}

\title{A Downward Collapse within the Polynomial 
Hierarchy}

\author{
Edith Hemaspaandra\protect\thanks{
Department of Mathematics,
Le Moyne College,
Syracuse, NY 13214, USA\@.
Supported in part 
by grant
NSF-INT-9513368/DAAD-315-PRO-fo-ab.  Work done in part while 
visiting 
Friedrich-Schiller-Universit\"at Jena.}
\and
Lane A. Hemaspaandra\thanks{
Department of Computer Science, 
University of Rochester, 
Rochester, NY 14627, USA\@.
Supported in part 
by grants NSF-CCR-9322513 and 
NSF-INT-9513368/DAAD-315-PRO-fo-ab.  Work done in part while 
visiting 
Friedrich-Schiller-Universit\"at Jena.}
\and 
Harald Hempel\thanks{
Inst.~f\"ur Informatik, 
Friedrich-Schiller-Universit\"at Jena, 
07743 Jena, Germany. 
Supported in part 
by grant
NSF-INT-9513368/DAAD-315-PRO-fo-ab.  Work done in part 
while visiting Le~Moyne College.}
}

\date{June 16, 1997}

\makeatletter
\def\@listI{\leftmargin\leftmargini \parsep 4.5pt plus 1pt minus 1pt\topsep
6pt plus 2pt minus 2pt \itemsep  2pt plus 2pt minus 1pt}

\let\@listi\@listI
\@listi
\makeatother

\begin{document}

\typeout{PLEASE NEVER DELETE COMMENTED OUT STUFF!! It is part of the paper's memory.  Cheers, Lane}
\typeout{PLEASE NEVER DELETE COMMENTED OUT STUFF!! It is part of the paper's memory.  Cheers, Lane}
\typeout{PLEASE NEVER DELETE COMMENTED OUT STUFF!! It is part of the paper's memory.  Cheers, Lane}

\typeout{WARNING:  BADNESS used to suppress reporting!  Beware!!}
\hbadness=3000%
\vbadness=10000 %

\bibliographystyle{alpha}

\setcounter{page}{1}

{\singlespacing\maketitle}

\begin{abstract}
Downward collapse (a.k.a.~upward separation) refers to cases where
the equality of two larger classes implies the equality of two
smaller classes.  
We provide an unqualified downward collapse result
completely within the polynomial hierarchy.  In particular, we prove
that, for $k > 2$, if $\psigkone = \psigktwo$ then $\sigmak = \pik = \ph$.
We extend this to obtain a more general downward collapse result.
\end{abstract}

\setcounter{page}{1}
\sloppy

\section{Introduction}
The theory of NP-completeness does not resolve the issue of whether~P
and~NP are equal.  However, it does unify the issues of whether
thousands of natural problems---the NP-complete problems---have
deterministic polynomial-time algorithms.  The study of downward
collapse is similar in spirit.  By proving downward collapses,
we seek to tie together central open issues regarding the computing
power of complexity classes.  For example, the 
main result of this paper shows that (for
$k>2$) the issue of whether the $k$th level of the polynomial
hierarchy is closed under complementation is identical to the issue of
whether two queries to this level give more power than one query 
to this level.

Informally, downward collapse (equivalent terms 
are ``downward translation of equality'' and ``upward separation'')
refers to cases in which
the collapse of larger classes implies the collapse of smaller 
classes (for background, see, e.g.,~\cite{all:j:lim,all-wil:j:downward}).
For example, $\npnp = \conpnp \Rightarrow \np = \conp$ would be a
(shocking, and inherently nonrelativizing~\cite{ko:j:exact}) 
downward collapse, the
``downward'' part referring to
the well-known fact that  
$\np \cup \conp
\subseteq
\npnp \cap \conpnp$.  

Downward collapse results are extremely rare, but there are 
some results in the literature that do have the general flavor of 
downward collapse.
Cases where the collapse of larger classes forces sparse sets (but 
perhaps not non-sparse sets) to fall out of smaller classes 
were found by Hartmanis, Immerman, and Sewelson 
(\cite{har-imm-sew:j:sparse}, see 
also~\cite{boo:j:tally})
and by
others (e.g., Rao, Rothe,
and Watanabe~\cite{rao-rot-wat:j:upward}, but in contrast
see also~\cite{hem-jha:j:defying}).
Existential cases have long been implicitly known (i.e., 
theorems such as ``If $\ph=\pspace$ then 
$(\exists k)\,[\ph=\sigmak]$''---note that here one can prove 
nothing about what value $k$ might have).  
Regarding probabilistic classes, 
Ko~\cite{ko:j:bpp}
proved that ``If $\rm NP \subseteq BPP$ then $\rm NP = R$,''
and
Babai, Fortnow, Nisan, and 
Wigderson~\cite{bab-for-nis-wig:j:publishable-proofs}
proved the striking result 
that ``If $\rm EH = E$ then $\p=\bpp$.'' 
Hemaspaandra, Rothe, and Wechsung have given 
an example involving degenerate
certificate schemes~\cite{hem-rot-wec:jtoappear:hard-certificates},
and examples due to 
Allender~\cite[Section~5]{all:coutdatedExceptForPUNCstuff:complexity-sparse}
and Hartmanis and Yesha~\cite[Section~4]{har-yes:j:computation} 
are known regarding 
circuit-related classes.\footnote{Note that we are not claiming 
that all the above examples from
the literature are totally
unqualified downward collapse results, but rather we are merely
stating that they have the strong general flavor of downward collapse.
In some cases, the results mentioned above do not fully witness what
one might hope for from the notion of ``downward.''  Ideally,
downward collapse results would be truly ``downward'' in the 
sense that they would be of the form ``If ${\cal A} = {\cal B}$ then
${\cal C} ={\cal D}$,'' where the classes are such that 
(a)~${\cal A}
\cap {\cal B} \supseteq {\cal C} \cup {\cal D}$ is a well-known
result, and (b)~it is not currently known that ${\cal A} \cap
{\cal B} = {\cal C} \cup {\cal D}$.  The downward collapses proven in
this paper do have this strong ``downward'' form.}

We provide an unqualified downward collapse result that is not
restricted to sparse or tally sets, whose conclusion does not contain
a variable that is not specified in its hypothesis, and that deals
with classes whose {\em ex ante\/} containments\footnote{%
\protect\singlespacing\protect\label{f:footnoteone}I.e., in the case of
Theorem~\protect\ref{t:main}, 
$\sigmak \cup \pik
\subseteq 
\psigkone \cap \psigktwo $ is well-known to be true (and most
researchers suspect that the inclusion is strict).}
are clear (and plausibly strict).
Namely, as is standard, let $\p^{{\cal C}[j]}$ denote the class of
languages computable by \p\ machines making at most $j$ queries to
some set from ${\cal C}$.  We prove that, for each $k > 2$, it holds
that $$\psigkone = \psigktwo \Rightarrow \sigmak = \pik = \ph.$$
(As just mentioned in footnote~\ref{f:footnoteone},
the classes in the hypothesis clearly have the property
that they contain both $\sigmak$ and $\pik$.)  
The best previously known results from the assumption 
$\psigkone = \psigktwo$ collapse the polynomial hierarchy
only to a level that contains 
$\sigmakplusone$ and 
$\pikplusone$~\cite{cha-kad:j:closer,bei-cha-ogi:j:difference-hierarchies}.

Our proof actually establishes a $\sigmak = \pik$ collapse from a
hypothesis that is even weaker than $\psigkone = \psigktwo$.  Namely,
we prove that, for $i<j<k$ and $i< k-2$, if one query each (in parallel)
to the
$i$th and $k$th levels of the polynomial hierarchy equals one query
each (in parallel) 
to the $j$th and $k$th levels of the polynomial hierarchy, then
$\sigmak = \pik = \ph$.
  
In the final section of the paper, we generalize from 1-versus-2
queries to $m$-versus-$(m+1)$ queries.  In particular, we show that
our main result is in fact a reflection of an even more general
downward collapse: If the truth-table hierarchy over $\sigmak$
collapses to its $m$th level, then the boolean hierarchy over
$\sigmak$ collapses one level further than one would expect.

\section{Simple Case}\label{s:simple}

Our proof works by
extracting advice internally and algorithmically, while 
holding down the number of quantifiers needed, within the 
framework of a so-called
``easy-hard'' argument.  Easy-hard arguments were introduced by
Kadin~\cite{kad:joutdatedbychangkadin:bh}, and 
were further used by 
Chang and Kadin~(\cite{cha-kad:j:closer}, see 
also~\cite{cha:thesis:boolean}) and 
Beigel, Chang, and Ogihara~\cite{bei-cha-ogi:j:difference-hierarchies}
(we follow the approach of Beigel, Chang, and 
Ogihara).

\begin{theorem}
\label{t:main}
For each $k>2$ it holds that:
$$\psigkone = \psigktwo \Rightarrow \sigmak = \pik = \ph.$$
\end{theorem}
Theorem~\ref{t:main} follows 
immediately\footnote{In particular, taking $i=0$ and
$j=k-1$ in Theorem~\protect\ref{t:np} yields a statement
that itself clearly implies Theorem~\protect\ref{t:main}.}
from Theorem~\ref{t:np}
below, which states that, for $i<j<k$ and $i< k-2$,
if one query each to the $i$th and $k$th levels of the
polynomial hierarchy equals one query each to the $j$th and $k$th
levels of the polynomial hierarchy, then $\sigmak = \pik = \ph$.

DPTM will refer to deterministic polynomial-time oracle Turing
machines, whose polynomial time upper-bounds are clearly clocked, and
are independent of their oracles.  We will also 
use the following definitions.

\begin{definition}
\begin{enumerate}
\item
Let $M^{(A, B)}$ denote DPTM $M$ making,
simultaneously (i.e., in a truth-table fashion), 
at most one query to oracle $A$ and 
at most one query to oracle $B$,
and let $$\p^{({\cal C},{\cal D})}=
\{L \subseteq \Sigma^* \condition (\exists C \in {\cal C})(\exists D \in 
{\cal D})(\exists {\rm DPTM}~M)[L=L(M^{(C,D)})]\}.$$
\item  (see~\cite{bei-cha-ogi:j:difference-hierarchies})~~$A \deltatilde 
B = \{ \pair{x, y} \condition x \in A \Leftrightarrow
y \not \in B\}.$
\end{enumerate}
\end{definition}

\begin{lemma}
\label{l:np}
Let $0 \leq i < k$, let
$\lpsigi$ be any set \manyone-complete for $\psigione$,
and let $\lsigk$ be any language \manyone-complete for $\sigmak$.
Then $\lpsigi \deltatilde \lsigk$ is \manyone-complete for 
$\psigik$.
\end{lemma}

\begin{proof}
Clearly $\lpsigi \deltatilde \lsigk$ is in
$\psigik$.
Regarding $\manyone$-hardness for 
$\psigik$,
let $L \in \psigik$ via transducer $M$, $\sigmai$ set $A$, and $\sigmak$
set $B$. Without loss of 
generality, on each input $x$, $M$ asks exactly one question $a_x$
to $A$, and one question $b_x$ to $B$.  Define sets $D$ and $E$
as follows:
\begin{itemize}
\item[]
$D = \{x \condition M^{(A,B)}$ accepts $x$ if $a_x$ is answered
correctly, and $b_x$ is answered ``no''$\}$.
\item[]
$E = \{x \condition  b_x \in B$ and the (one-variable)
truth-table with respect to $b_x$
of $M^{(A,B)}$  on input $x$ induced by the correct
answer to $a_x$ 
is neither
``always accept'' nor ``always reject''$\}$.
\end{itemize}
Note that $D \in \psigione$, and that $E \in \sigmak$, since $i < k$.
But $L \manyone D \deltatilde E$ via the reduction $f(x) =
\pair{x, x}$.  So clearly $L \manyone \lpsigi \deltatilde \lsigk$,
via the reduction 
$\widehat{f}(x) =
\pair{f'(x), f''(x)}$, where $f'$ and $f''$ are, respectively, reductions
{}from $D$ to 
$\lpsigi$ and from
$E$ to $\lsigk$.\qquad\end{proof}

Theorem~\ref{t:np} contains the following two technical advances.
First, it internally extracts information in a way
that saves a quantifier.
(In contrast, the earliest easy-hard arguments in the literature
merely ensure that $\sigmak \subseteq \pik/\poly$ and from that infer
a weak polynomial
hierarchy collapse.  Even the interesting recent strengthenings of the
argument~\cite{bei-cha-ogi:j:difference-hierarchies} still, under the
hypothesis of Theorem~\ref{t:np}, conclude only a collapse of the
polynomial hierarchy to a level a bit above $\Sigma^p_{k+1}$.)  The
second advance is that previous easy-hard arguments seek to determine
whether there exists a hard string for a length or not.  Then they use
the fact that if there is not a hard string, all strings (at the
length) are easy.  In contrast, 
we {\em never\/} search for a hard string; rather,
we use the fact that the input itself (which we do not have to search
for as, after all, it is our input) is either easy or hard.  So we
check whether the input is easy, and if so we can use it as an easy
string, and if not, it must be a hard string so we can use it that
way.  This innovation is important in that it allows
Theorem~\ref{t:main} to apply for all $k>2$---as opposed to merely
applying for all $k>3$, which is what we would get without this
innovation.  (Following a referee's suggestion, we mention that
during a first traversal the reader may wish to consider just 
the $i=0$ and $j=1$ special case of 
Theorem~\ref{t:np} and its proof, as this provides a restricted 
version that is easier to read.)

\begin{theorem}\label{t:np}
Let $0 \leq i < j < k$ and $i < k-2$.
If $\psigik=\psigjk$ then $\sigmak = \pik = \ph.$
\end{theorem}

\begin{proof}
Suppose $\psigik=\psigjk$.
Let $\lpsigi$, $\lpsigipone$, and $\lsigk$ be \manyone-complete 
for $\psigione$,
$\psigipone$, and $\sigmak$, respectively;  such sets exist.
{}From Lemma~\ref{l:np} it follows that 
$\lpsigi \deltatilde \lsigk$ is \manyone-complete for $\psigik$.
Since (as $i<j$) 
$\lpsigipone \deltatilde \lsigk \in \psigjk$, and by assumption
$\psigjk = \psigik$,
there exists a polynomial-time many-one reduction $h$ from
$\lpsigipone \deltatilde \lsigk$ to
$\lpsigi \deltatilde \lsigk$.
So, for all $x_1, x_2 \in \sigmastar$:
if $h(\pair{x_1, x_2}) = \pair{y_1, y_2}$, 
then
$(x_1 \in \lpsigipone \Leftrightarrow x_2 \not\in \lsigk)$ 
if and only if
$(y_1 \in \lpsigi \Leftrightarrow y_2 \not\in \lsigk)$.
Equivalently,
for all $x_1, x_2 \in \sigmastar$:
\begin{quotation}
\noindent
{\bf Fact~1:} \\
if $h(\pair{x_1, x_2}) = \pair{y_1, y_2}$, \\ 
then
$$(x_1 \in \lpsigipone \Leftrightarrow x_2 \in \lsigk)
\mbox{ if and only if }
(y_1 \in \lpsigi \Leftrightarrow y_2 \in \lsigk).$$
\end{quotation}

We can use $h$ to recognize some of $\overline{\lsigk}$ by a $\sigmak$
algorithm.  The definitions of easy and hard used in this paper follow
the easy and hard concepts used by 
Kadin~\cite{kad:joutdatedbychangkadin:bh}, 
Chang and Kadin~(\cite{cha-kad:j:closer}, see 
also~\cite{cha:thesis:boolean}), and 
Beigel, Chang, and Ogihara~\cite{bei-cha-ogi:j:difference-hierarchies},
modified as needed for our goals.
In particular,
we say that a string $x$ is {\em
easy for length $n$\/} if there exists a string $x_1$ such that
$|x_1| \leq n$ and $(x_1 \in \lpsigipone
\Leftrightarrow y_1 \not \in \lpsigi)$ where
$h(\pair{x_1, x}) = \pair{y_1, y_2}$.

Let $p$ be a fixed polynomial, which will be exactly 
specified later in the proof.
We have the following $\sigmak$ algorithm 
to test whether $x \in \overline{\lsigk}$ in the
case that (our input) $x$ is an easy string for $p(|x|)$.
On input $x$, guess $x_1$ with $|x_1| \leq p(|x|)$, let
$h(\pair{x_1, x}) = \pair{y_1, y_2}$,
and accept if and only if
$(x_1 \in \lpsigipone \Leftrightarrow y_1 \not \in \lpsigi)$  and
$y_2 \in \lsigk$.  In light of Fact~1 above, it is clear that this 
is correct.

We say that $x$ is {\em hard for length $n$\/} if 
$|x| \leq n$ and $x$ is not easy for length $n$, i.e., if
$|x| \leq n$ and for all $x_1$ with $|x_1| \leq n$, $(x_1 \in \lpsigipone
\Leftrightarrow y_1 \in \lpsigi)$, where
$h(\pair{x_1, x}) = \pair{y_1, y_2}$.

If $x$ is a hard string for length $n$, then $x$ induces a
many-one reduction from 
${\left(\lpsigipone\right)}^{\leq n}$ to $\lpsigi$,
namely, $f(x_1) = y_1$, where $h(\pair{x_1, x}) = \pair{y_1, y_2}$.
Note that $f$ is computable in time 
polynomial in $\max(n,|x_1|)$.

We can use hard strings to obtain a $\sigmak$ algorithm 
for $\overline{\lsigk}$.
Let $M$ be a $\Pi_{k-i-1}^p$ machine such that $M$ with oracle
$\lpsigipone$ recognizes $\overline{\lsigk}$.
Let the run-time of $M$ be bounded by polynomial $p$,
which without loss of generality satisfies
$(\forall 
\widehat{m} \geq 0)[p(\widehat{m}+1) > 
p(\widehat{m}) > 0]$ (as promised above, we have
now specified $p$).
Then $${\left(\overline{\lsigk}\right)}^{= n} = 
L 
( 
{
M^{   {\left( \lpsigipone
\right)}^{\leq p(n)}
}
} 
) 
{}^{= n}.$$
If there exists a hard string for length $p(n)$, then this hard string
induces a reduction from 
${\left(\lpsigipone\right)}^{\leq p(n)}$ to $\lpsigi$.
Thus, with
any hard string for length $p(n)$ in hand, call it $w_n$, 
$\widehat{M}$ with oracle $\lpsigi$
recognizes $\overline{\lsigk}$ for strings of length $n$, 
where $\widehat{M}$ is the machine that simulates $M$ but replaces
each query to $q$ by the first component of $h(\pair{q,w_n})$.
It follows that if there exists a hard string for length $p(n)$, then
this string induces a $\Pi_{k-1}^p$ algorithm for 
${\left(\overline{\lsigk}\right)}^{=n}$, and therefore certainly
a $\sigmak$ algorithm for
${\left(\overline{\lsigk}\right)}^{=n}$.

However, now we have an $\np^{\Sigma_{k-1}^p} =\sigmak$ algorithm 
for 
$\overline{\lsigk}$:  On input $x$, the NP base machine of 
$\np^{\Sigma_{k-1}^p}$ executes the following algorithm:
\begin{enumerate}
\item Using its 
$\Sigma_{k-1}^p$ oracle, it deterministically 
determines whether the input $x$ is an easy string for 
length $p(|x|)$.  This can be done, as checking whether 
the input is an easy string for length $p(|x|)$ can be done 
by one query to $\Sigma_{i+2}^p$, and $i+2 \leq k-1$ by 
our $i < k-2$ hypothesis.
\item If the previous step determined that the input is not 
an easy string, then the input must be a hard string
for length $p(|x|)$.  
So simulate the $\sigmak$ algorithm induced by this hard string
(i.e., the input $x$ itself) on input $x$ (via our NP
machine itself simulating the base level of the 
$\sigmak$ algorithm and using the NP machine's oracle to 
simulate the oracle queries made by the base level NP machine 
of the 
$\sigmak$ algorithm being simulated).
\item If the first step determined 
that the input $x$ is easy for length $p(|x|)$, then our NP
machine
simulates (using itself and its oracle) 
the $\sigmak$ algorithm for easy strings on input $x$.
\end{enumerate}
We need one brief technical comment.  The 
$\Sigma_{k-1}^p$ oracle in the above algorithm is being used
for a number of different sets.  However, as 
$\Sigma_{k-1}^p$ is closed under disjoint 
union, this presents no 
problem as we can use the disjoint union of the sets, 
while modifying the queries so they address the 
appropriate part of the disjoint union.

Since $\overline{\lsigk}$ is complete for $\pik$,
it follows that  $\sigmak = \pik = \ph$.
\qquad\end{proof}

We conclude this section with three remarks.  First, if one is 
fond of the truth-table version of bounded 
query hierarchies, one can certainly replace the
hypothesis of Theorem~\ref{t:main} with
$\p_{1 \mbox{\scriptsize{}-tt}}^{\sigmak} =
\p_{2 \mbox{\scriptsize{}-tt}}^{\sigmak}$ (both as this is 
an equivalent hypothesis, and as it in any case clearly follows
{}from Theorem~\ref{t:np}).   Indeed, one can equally well
replace the hypothesis of Theorem~\ref{t:main} with the 
even weaker-looking hypothesis\footnote{%
\protect\singlespacing Where ${\rm DIFF}_2(\calc) =_{\rm{}def}
\{ L \condition (\exists L_1 \in \calc)(\exists L_2\in \calc)
[L = L_1-L_2]\}$, and ${\rm co}\calc =_{\rm{}def} \{L \condition
\overline{L} \in \calc\}$ (see 
Definition~\protect\ref{d:diff} for background).}
$\p^{\sigmak [1]} = {\rm DIFF}_2(\sigmak)$ (as this hypothesis
is also in fact equivalent to the hypothesis of 
Theorem~\ref{t:main}---just note that if
$\p^{\sigmak [1]} = {\rm DIFF}_2(\sigmak)$ then 
${\rm DIFF}_2(\sigmak)$ is closed under complementation and 
thus equals the boolean hierarchy over 
$\sigmak$, see~\cite{cai-gun-har-hem-sew-wag-wec:j:bh1}, 
and so in particular we then have 
$\p^{\sigmak [1]} = {\rm DIFF}_2(\sigmak) = \p^{\sigmak [2]}$).

Of course, the two equivalences just 
mentioned---$\psigkone = \psigktwo 
\Leftrightarrow
\p_{1\hbox{-}\rm{}tt}^{\sigmak} =  
\p_{2\hbox{-}\rm{}tt}^{\sigmak}
\Leftrightarrow
\p^{\sigmak [1]} = {\rm DIFF}_2(\sigmak)$---are well-known.
However, Theorem~\ref{t:np} is sufficiently strong that it 
creates an equivalence that is quite new, and somewhat 
surprising.  We state it below as Corollary~\ref{c:amazing}.
\begin{theorem}\label{t:tricky-t}
For each $k>2$ it holds that:
$$\psigkone = {\rm DIFF}_2(\sigmak) \cap {\rm coDIFF}_2(\sigmak) 
 \Rightarrow \sigmak = \pik = \ph.$$
\end{theorem}
\begin{proof}
Let $A \triangle B =_{\rm def} (A-B) \cup (B-A)$.
Recalling that $k>2$, it is not 
hard to see that 
$\psigjustnpk \subseteq {\rm DIFF}_2(\sigmak)$.   
In particular, this holds due 
to Lemma~\ref{l:np}, in light of the 
facts 
that 
(i)~${\rm DIFF}_2(\sigmak) = \{ L \condition
(\exists L_1 \in \sigmak)(\exists L_2 \in \sigmak)\allowbreak
[ L = L_1 \triangle L_2]\}$ (due
to K\"obler, Sch\"oning, and Wagner~\cite{koe-sch-wag:j:diff}---see
the discussion just before Theorem~\ref{t:twd}), 
and 
(ii)~$A \deltatilde B = 
\{\pair{x, y} \condition x\in 
A\} 
\: \triangle \: 
\{\pair{x, y} \condition y\in 
B\}$.  So, since
$\psigjustnpk$ is closed under complementation, we have
$\psigkone \subseteq \psigjustnpk \subseteq {\rm DIFF}_2(\sigmak) \cap 
{\rm coDIFF}_2(\sigmak)$.  However, this says, under the 
hypothesis of the theorem, that 
$\psigkone = \psigjustnpk$, which itself, by
Theorem~\ref{t:np}, implies that $\sigmak = \pik = \ph$.\qquad\end{proof}

\begin{corollary}\label{c:amazing}
For each $k>2$ it holds that:
$$\psigkone = {\rm DIFF}_2(\sigmak) \cap {\rm coDIFF}_2(\sigmak) 
\Leftrightarrow 
\psigkone = {\rm DIFF}_2(\sigmak).$$
\end{corollary}
Our second remark is that Theorem~\ref{t:main}
implies that, for $k>2$, if the bounded query hierarchy over 
$\sigmak$ collapses to its $\psigkone$ level, then the 
bounded query hierarchy over $\sigmak$ equals the polynomial
hierarchy (this provides a partial answer to the issue of 
whether, when a bounded query hierarchy collapses, the polynomial
hierarchy necessarily collapses to it, 
see~\cite[Problem~4]{hem-ram-zim:j:worlds-to-die-for}).

Third, in Lemma~\ref{l:np} and Theorem~\ref{t:np} we speak of classes of 
the form $\psigij$, $i\neq j$.  It would be very natural 
to reason as follows:  ``$\psigij$, $i\neq j$, must 
equal $\p^{\Sigma^p_{\max(i,j)}[1]}$, as 
$\Sigma^p_{\max(i,j)}$ can easily solve any 
$\Sigma^p_{\min(i,j)}$ query ``strongly'' using the 
$\Sigma^p_{\max(i,j)-1}$ oracle of its base NP machine
and thus the hypothesis of Theorem~\ref{t:np} is trivially
satisfied and so you in fact are claiming to prove, unconditionally,
that $\ph = \Sigma_3^p$.  This reasoning, though tempting, is wrong
for the following somewhat subtle reason.  Though it is 
true that, for example, $\np^{\sigmaq}$ can solve any $\sigmaq$
query
and then can tackle any $\Sigma^p_{q+1}$ query, it does not 
follow that 
$\p^{(\Sigma^p_{q+1} , \sigmaq)} = \p^{ \Sigma^p_{q+1}[1]}$.
The problem is that the answer to the $\sigmaq$ query may 
{\em change the truth-table\/} the P transducer uses to 
evaluate the answer of the $\Sigma^p_{q+1}$ query.  

We mention that Buhrman and
Fortnow~\cite{buh-for:t:two-queries},
building on and 
extending our proof technique, have very recently obtained the
$k=2$ analog of Theorem~\ref{t:main}.  They also prove that there are
relativized worlds in which the $k=1$ analog of Theorem~\ref{t:main}
fails.  On the other hand, if one changes
Theorem~\ref{t:main}'s left-hand-side classes to 
function classes, then the $k=1$ analog
of the resulting claim does
hold due to 
Krentel (see~\cite[Theorem~4.2]{kre:j:optimization}):
$\rm FP^{NP[1]} = FP^{NP[2]} \implies P = PH$.
Also, very recent work of Hemaspaandra, Hemaspaandra, and 
Hempel~\cite{ehem-hem-hem:t:translating-equality}, building on the 
techniques of the present paper and those of Buhrman and 
Fortnow~\cite{buh-for:t:two-queries}, has established the $k=2$ analog of 
Theorem~\ref{t:genmain}.

\section{General Case}

We now generalize the results of Section~\ref{s:simple} to the case of
$m$-truth-table reductions.  Though the results of this section are
stronger than those of Section~\ref{s:simple}, the proofs are somewhat
more involved, and thus we suggest the reader first read
Section~\ref{s:simple}.  

For clarity, we now describe the two key differences between 
the proofs in this section and those of 
Section~\ref{s:simple}.  (1)~The completeness claims of
Section~\ref{s:simple} were simpler.  Here, we now need 
Lemma~\ref{l:gennp}, which extends
\cite[Lemma~8]{bei-cha-ogi:j:difference-hierarchies} with the 
trick of splitting a truth-table along a simple query's 
dimension in such a way that the induced one-dimension-lower
truth-tables cause no problems.  (2)~The proof of
Theorem~\ref{t:gennp} is quite analogous to the proof 
of Theorem~\ref{t:np}, except (i)~it is a bit harder to understand
as one continuously has to parse the deeply nested set differences
caused by the fact that we are now working in the difference 
hierarchy, and (ii)~the ``input is an easy string'' simulation
is changed to account for a new problem, namely, that in the 
boolean hierarchy one models each language by 
a {\em collection\/} of machines 
(mimicking the nested difference structure of boolean 
hierarchy languages) and thus it is 
hard to ensure that these machines, when guessing an object,
necessarily guess the same object (we solve this 
coordination problem by 
forcing them to each guess a lexicographically extreme object,
and we argue that this can be accomplished within the computational
power available).

The difference hierarchy was introduced by Cai et
al.~\cite{cai-gun-har-hem-sew-wag-wec:j:bh1,cai-gun-har-hem-sew-wag-wec:j:bh2}
and is defined below.  Cai et al.~studied the case $\calc = \np$,
but a number of other cases have since been 
studied~\cite{bru-jos-you:j:strong,bei-cha-ogi:j:difference-hierarchies,hem-rot:jtoappear:boolean}.  

\begin{definition}\label{d:diff}
Let $\cal C$ be any complexity class.
\begin{enumerate}
\item $ {\rm DIFF}_1(\calc) = \calc$.
\item For any $k \geq 1$, ${\rm DIFF}_{k+1}(\calc) =
\{ L \condition  (\exists L_1 \in \calc)(\exists L_2 \in 
{\rm DIFF}_k(\calc))[ L = L_1 - L_2]\}$.
\item For any $k \geq 1$, $ {\rm coDIFF}_k(\calc) = 
\{ L \condition \overline{L} \in 
{\rm DIFF}_k(\calc)\}$.
\end{enumerate}
\end{definition}

Note in particular that
$$ {\rm DIFF}_m(\sigmak) \cup
{\rm coDIFF}_m(\sigmak) 
\subseteq \p_{m \hbox{-}{\rm tt}}^{\sigmak} \subseteq
{\rm DIFF}_{m+1}(\sigmak) \cap
{\rm coDIFF}_{m+1}(\sigmak).$$

\begin{theorem}
\label{t:genmain}
For each $m > 0$ and each $k>2$ it holds that:
$$\psigkmtt = \psigkmponett \Rightarrow \diffmsigk = \codiffmsigk.$$
\end{theorem}
Theorem~\ref{t:main} is the $m = 1$ case 
of Theorem~\ref{t:genmain} (except the former is stated in terms 
of Turing access).
Theorem~\ref{t:genmain} follows immediately from Theorem~\ref{t:gennp}
below, which states that, for $i<j<k$ and $i< k-2$, if one query to
the $i$th and $m$ queries to the $k$th levels of the polynomial
hierarchy equals one query to the $j$th and $m$ queries to the $k$th
levels of the polynomial hierarchy, then $\diffmsigk = \codiffmsigk$.
Note, of course, that by Beigel, Chang, and 
Ogihara~\cite{bei-cha-ogi:j:difference-hierarchies} 
the conclusion of
Theorem~\ref{t:genmain} implies a collapse of the 
polynomial hierarchy.  In
particular, 
via~\cite[Theorem~10]{bei-cha-ogi:j:difference-hierarchies},
Theorem~\ref{t:genmain} implies that, for each $m \geq 0$
and each $k>2$, it holds that: If $\psigkmtt = \psigkmponett$ then the
polynomial hierarchy can be solved by a P machine that makes $m-1$
truth-table queries to $\Sigma^p_{k+1}$, and that in addition is
allowed unbounded queries to $\Sigma^p_{k}$.  This polynomial
hierarchy collapse is about one level lower in the difference
hierarchy over $\Sigma_{k+1}^p$ than one could conclude from previous
papers, in particular, from Beigel, Chang, and
Ogihara.
In fact, one can claim a bit more.  The {\em proof\/} of 
\cite[Theorem~10]{bei-cha-ogi:j:difference-hierarchies}
in fact proves the following:  $\diffmsigk = \codiffmsigk
\Rightarrow \ph = 
{\p^{(\sigmak , \Sigma_{k+1}^p)}_{1,m-1 \hbox{-}{\rm tt}}}$.
Thus, in light of Theorem~\ref{t:genmain}, we have 
the following corollary.
\begin{corollary}
\label{c:qrs-cor}
For each $m \geq 0$ and each $k>2$ it holds that
$\psigkmtt = \psigkmponett \Rightarrow 
\ph = 
{\p^{(\sigmak , \Sigma_{k+1}^p)}_{1,m-1 \hbox{-}{\rm tt}}}$.
\end{corollary}

The following definition will be useful.

\begin{definition}
Let $M^{(A, B)}_{a,b\hbox{-}{\rm tt}}$ denote DPTM $M$ making,
simultaneously (i.e., all $a+b$ queries are made at 
the same time, in the standard truth-table fashion), 
at most $a$ queries to oracle $A$ and 
at most $b$ queries to oracle $B$,
and let $$\p^{({\cal C},{\cal D})}_{a,b\hbox{-}{\rm tt}}=
\{L \subseteq \Sigma^* \condition (\exists C \in {\cal C})(\exists D \in 
{\cal D})(\exists {\rm DPTM}~M)[L=L(M^{(C,D)}_{a,b\hbox{-}{\rm tt}})]\}.$$
\end{definition}

\begin{lemma}
\label{l:gennp}
Let $m >0$, let $0 \leq i < k$, let
$\lpsigi$ be any set \manyone-complete for $\psigione$,
and let $\ldiffmsigk$ be any language \manyone-complete for $\diffmsigk$.
Then $\lpsigi \deltatilde \ldiffmsigk$ is \manyone-complete for 
$\psigimk$.
\end{lemma}

Lemma~\ref{l:gennp} does not require proof, as it is a use of
the standard mind-change technique, and is analogous
to~\cite[Lemma~8]{bei-cha-ogi:j:difference-hierarchies}, with one key
twist that we now discuss.  
Assume, without loss of generality, that we focus on 
$\psigimk$ machines that always make exactly $m+1$ queries.
Regarding any such machine accepting a
set complete for the class
$\psigimk$ of Lemma~\ref{l:gennp}, we have on each input
a truth-table with $m+1$
variables.  Note that if one knows the answer to the one $\sigmai$
query, then this induces a truth-table on $m$ variables; however, note
also that the two $m$-variable truth-tables (one corresponding to a
``yes'' answer to the $\sigmai$ query and the other to a ``no''
answer) may differ sharply.  Regarding $\lpsigi \deltatilde
\ldiffmsigk$, we use $\lpsigi$ to determine whether the $m$-variable
truth-table induced by the true answer to the one $\sigmai$ query
accepts or not when all the $\sigmak$ queries get the answer no.  This
use is analogous
to~\cite[Lemma~8]{bei-cha-ogi:j:difference-hierarchies}.  The new
twist is the action of the $\ldiffmsigk$ part of $\lpsigi \deltatilde
\ldiffmsigk$.  We use this, just as in
\cite[Lemma~8]{bei-cha-ogi:j:difference-hierarchies}, to find whether
or not we are in an odd mind-change region {\em but now with respect
to the $m$-variable truth-table induced by the true answer to the one
$\sigmai$ query}.  Crucially, this still is a $\diffmsigk$ issue as,
since $i<k$, a $\sigmak$ machine can first on its own (by its base NP
machine making one deterministic query to its $\Sigma_{k-1}^p$ oracle)
determine the true answer to the one $\sigmai$ query, and thus the
machine can easily know which of the two $m$-variable truth-table
cases it is in, and thus it plays its standard part in determining if
the mind-change region of the $m$ true answers to 
the $\sigmak$ queries fall in an odd mind-change
region {\em with respect to the correct $m$-variable truth-table}.

\begin{theorem}\label{t:gennp}
Let $m >  0$,  $0 \leq i < j < k$ and $i < k-2$.
If $\psigimk=\psigjmk$ then $\diffmsigk = \codiffmsigk$.
\end{theorem}

\begin{proof}  
Suppose $\psigimk=\psigjmk$.
Let $\lpsigi$, $\lpsigipone$, and $\ldiffmsigk$ be \manyone-complete for
$\psigione$,
$\psigipone$, and $\diffmsigk$, respectively;  such languages
exist, e.g., via the standard canonical complete set 
constructions using enumerations of clocked machines.
{}From Lemma~\ref{l:gennp} it follows that 
$\lpsigi \deltatilde \ldiffmsigk$ is \manyone-complete for $\psigimk$.
Since (as $i<j$) 
$\lpsigipone \deltatilde \ldiffmsigk \in \psigjmk$, and by assumption
$\psigjmk = \psigimk$,
there exists a polynomial-time many-one reduction $h$ from
$\lpsigipone \deltatilde \ldiffmsigk$ to
$\lpsigi \deltatilde \ldiffmsigk$.
So, for all $x_1, x_2 \in \sigmastar$:
\begin{quotation}
\noindent
if $h(\pair{x_1, x_2}) = \pair{y_1, y_2}$, \\ 
then
$$(x_1 \in \lpsigipone \Leftrightarrow x_2 \in \ldiffmsigk)
\mbox{ if and only if }
(y_1 \in \lpsigi \Leftrightarrow y_2 \in \ldiffmsigk).$$
\end{quotation}

We can use $h$ to recognize some of $\overline{\ldiffmsigk}$ by a $\diffmsigk$
algorithm. 
In particular, 
we say that a string $x$ is {\em
easy for length $n$\/} if there exists a string $x_1$ such that
$|x_1| \leq n$ and $(x_1 \in \lpsigipone
\Leftrightarrow y_1 \not \in \lpsigi)$ where
$h(\pair{x_1, x}) = \pair{y_1, y_2}$.

Let $p$ be a fixed polynomial, which will be exactly 
specified later in the proof.
We have the following algorithm 
to test whether $x \in \overline{\ldiffmsigk}$ in the
case that (our input) $x$ is an easy string for $p(|x|)$.
On input $x$, guess $x_1$ with $|x_1| \leq p(|x|)$, let
$h(\pair{x_1, x}) = \pair{y_1, y_2}$,
and accept if and only if
$(x_1 \in \lpsigipone \Leftrightarrow y_1 \not \in \lpsigi)$  and
$y_2 \in \ldiffmsigk$.  
This algorithm is not necessarily a $\diffmsigk$ algorithm,
but it does inspire the following
$\diffmsigk$ algorithm
to test whether $x \in \overline{\ldiffmsigk}$ in the
case that $x$ is an easy string for $p(|x|)$.
Let $L_1, L_2, \cdots, L_m$ be languages in $\Sigma^p_k$ such that
$\ldiffmsigk = L_1 - (L_2 - (L_3 - \cdots  (L_{m-1} - L_m) \cdots))$.
Then $x \in \overline{\ldiffmsigk}$ if and only if
$x \in L'_1 - (L'_2 - (L'_3 - \cdots  (L'_{m-1} - L'_m) \cdots))$,
where $L_r'$ is computed as follows: 
On input $x$, guess $x_1$ with $|x_1| \leq p(|x|)$, let
$h(\pair{x_1, x}) = \pair{y_1, y_2}$,
and accept if and only if
(a)~$(x_1 \in \lpsigipone \Leftrightarrow y_1 \not \in \lpsigi)$, and
(b)~$(\forall z <_{lex} x_1)
[z \in \lpsigipone \Leftrightarrow w_1 \in \lpsigi]$, where
$h(\pair{z, x}) = \pair{w_1, w_2}$, and
(c)~$y_2 \in L_r$.

Since $i+2 < k$, $L'_r \in \sigmak$, and thus our
algorithm is in $\diffmsigk$.  Note that condition~(b) has no
analog in the proof of Theorem~\ref{t:np}.  We need this 
extra condition here as otherwise the different $L'_r$ might 
latch onto {\em different\/} strings $x_1$ and this would 
cause unpredictable behavior (as different $x_1$s would
create different $y_2$s).

We say that $x$ is {\em hard for length $n$\/} if 
$|x| \leq n$ and $x$ is not easy for length $n$, i.e., if
$|x| \leq n$ and for all $x_1$ with $|x_1| \leq n$, $(x_1 \in \lpsigipone
\Leftrightarrow y_1 \in \lpsigi)$, where
$h(\pair{x_1, x}) = \pair{y_1, y_2}$.

If $x$ is a hard string for length $n$, then $x$ induces a
many-one reduction from 
${\left(\lpsigipone\right)}^{\leq n}$ to $\lpsigi$,
namely, $f(x_1) = y_1$, where $h(\pair{x_1, x}) = \pair{y_1, y_2}$.
Note that $f$ is computable in time 
polynomial in $\max(n,|x_1|)$.

We can use hard strings to obtain a $\diff_m(\Sigma^p_{k-1})$ algorithm 
for $\ldiffmsigk$, and thus
(since $\diff_m(\Sigma^p_{k-1}) \subseteq \p^{\Sigma^p_{k-1}}
\subseteq \sigmak \cap \pik$) 
certainly a $\diffmsigk$ algorithm for
$\overline{\ldiffmsigk}$.
Again, let $L_1, L_2, \cdots, L_m$ be languages in $\Sigma^p_k$ such that
$\ldiffmsigk = L_1 - (L_2 - (L_3 - \cdots  (L_{m-1} - L_m) \cdots))$.
For all $1 \leq r \leq m$, let $M_r$ be a
$\Sigma_{k-i-1}^p$ machine such that
$L_r  =  L  ( {M_r^{Y}} )$,
where $Y = 
\lpsigipone$.
Let the run-time of all $M_r$s be bounded by polynomial $p$,
which without loss of generality satisfies
$(\forall 
\widehat{m} \geq 0)[p(\widehat{m}+1) > 
p(\widehat{m}) > 0]$ (as promised above, we have
now specified $p$).
Then  for all $1 \leq r \leq m$,
$${\left({L_r}\right)}^{= n} = 
L 
( 
{
M_r^{\left( {Y}^{\leq p(n)}\right)
 }
} 
) 
{}^{= n},$$
where $Y = \lpsigipone$.
If there exists a hard string for length $p(n)$, then this hard string
induces a reduction from 
${\left(\lpsigipone\right)}^{\leq p(n)}$ to $\lpsigi$.
Thus, with
any hard string for length $p(n)$ in hand, call it $w_n$, 
$\widehat{M_r}$ with oracle $\lpsigi$
recognizes $L_r$ for strings of length $n$, 
where $\widehat{M_r}$ is the machine that simulates $M_r$ but replaces
each query to $q$ by the first component of $h(\pair{q,w_n})$.
It follows that if there exists a hard string for length $p(n)$, then
this string induces a $\diff_m({\Sigma_{k-1}^p})$ algorithm for 
${\left({\ldiffmsigk}\right)}^{=n}$, and therefore certainly
a $\diffmsigk$ algorithm for
${\left(\overline{\ldiffmsigk}\right)}^{=n}$.
It follows that there exist $m$ $\sigmak$ sets, say, $\widehat{L_r}$ for
$1 \leq r \leq m$, such that the following holds:
For all $x$, if $x$ (functioning as $w_{|x|}$ above)
is a hard string for length $p(|x|)$, then
$x \in \overline{\ldiffmsigk}$ if and only if
$x \in \widehat{L_1} - (\widehat{L_2} - (\widehat{L_3} - \cdots 
(\widehat{L_{m-1}} - \widehat{L_m}) \cdots))$.

However, now we have an outright $\diffmsigk$ algorithm 
for $\overline{\ldiffmsigk}$: For $1 \leq r \leq m$ define a
$\np^{\Sigma_{k-1}^p}$  machine  $N_r$ as follows:
On input $x$, the NP base machine of $N_r$ 
executes the following algorithm:
\begin{enumerate}
\item Using its 
$\Sigma_{k-1}^p$ oracle, it deterministically 
determines whether the input $x$ is an easy string for 
length $p(|x|)$.  This can be done, as checking whether 
the input is an easy string for length $p(|x|)$ can be done 
by one query to $\Sigma_{i+2}^p$, and $i+2 \leq k-1$ by 
our $i < k-2$ hypothesis.
\item If the previous step determined that the input is not 
an easy string, then the input must be a hard string
for length $p(|x|)$.  
So simulate the $\sigmak$ algorithm for $\widehat{L_r}$
induced by this hard string
(i.e., the input $x$ itself) on input $x$ (via our NP
machine itself simulating the base level of the 
$\sigmak$ algorithm and using the NP machine's oracle to 
simulate the oracle queries made by the base level NP machine 
of the 
$\sigmak$ algorithm being simulated).
\item If the first step determined 
that the input $x$ is easy for length $p(|x|)$, then our NP
machine
simulates (using itself and its oracle) 
the $\sigmak$ algorithm for  $L'_r$ on input $x$.
\end{enumerate}
It follows that  for all $x$, $x  
\in \overline{\ldiffmsigk}$ if and only if
$x \in L(N_1) - (L(N_2) - (L(N_3) - \cdots 
(L(N_{m-1}) - L(N_m)) \cdots))$.
Since $\overline{\ldiffmsigk}$ is complete for $\codiffmsigk$,
it follows that  $\diffmsigk = \codiffmsigk$.
\qquad\end{proof}

Finally, remark that we have analogs of 
Theorem~\ref{t:tricky-t}
and Corollary~\ref{c:amazing}.  The proof is
analogous to that of 
Theorem~\ref{t:tricky-t};  one just uses 
$\p_{1,m\hbox{-}\rm{}tt}^{(\np,\sigmak)}$ in the way 
$\psigjustnpk$ was used in that proof, and again
invokes the relation
between
the difference and symmetric difference hierarchies
(namely that ${\rm DIFF}_j(\sigmak)$ is exactly the 
class of sets $L$ that for some $L_1,\cdots,L_j\in\sigmak$
satisfy $L = L_1 \, \triangle \, \cdots \, \triangle \,
L_j$; this well-known equality is due 
to \cite[Section~3]{koe-sch-wag:j:diff} in light of 
the standard equalities regarding boolean
hierarchies (see~\cite[Section~2.1]{cai-gun-har-hem-sew-wag-wec:j:bh1});
though both 
\cite{koe-sch-wag:j:diff,cai-gun-har-hem-sew-wag-wec:j:bh1}
focus mostly on the $k=1$ case, it is 
standard~\cite{wec:c:bh:ormaybe:wech:only:is:right,ber-bru-jos-sit-you:c:gen}
that
the equalities in fact hold for any class closed under union and 
intersection and containing $\emptyset$ and $\sigmastar$).
\begin{theorem}\label{t:twd}
Let $m \geq 0$ and $k>2$.
If 
$\p_{m\hbox{-}\rm{}tt}^{\sigmak} = 
{\rm DIFF}_{m+1}(\sigmak) \cap
{\rm coDIFF}_{m+1}(\sigmak)$ 
then $\diffmsigk = \codiffmsigk$.
\end{theorem}

\begin{corollary}
For each $k>2$ and $m\geq 0$, it holds that:
$$\p_{m\hbox{-}\rm{}tt}^{\sigmak} = 
{\rm DIFF}_{m+1}(\sigmak) \cap
{\rm coDIFF}_{m+1}(\sigmak) 
\Leftrightarrow
\p_{m\hbox{-}\rm{}tt}^{\sigmak} = 
{\rm DIFF}_{m+1}(\sigmak).$$
\end{corollary}

{\samepage
\begin{center}
{\bf Acknowledgments}
\end{center}
\nopagebreak
\indent
The first two authors
thank Gerd Wechsung's research group for its
very kind hospitality during the visit when this 
research was performed.  The authors are grateful to 
two anonymous referees, Lance Fortnow, and J\"org
Rothe for helpful comments and suggestions.

}%

\singlespacing

\bibliography{gry}
\end{document}